# Dynamics of heterodimeric kinesins and cooperation of kinesins

Ping Xie, Shuo-Xing Dou, and Peng-Ye Wang

*Laboratory of Soft Matter Physics, Institute of Physics, Chinese Academy of Sciences,*
*P .O. Box, 603, Beijing 100080, China*
(*E-mail: pxie@aphy.iphy.ac.cn, sxdou@aphy.iphy.ac.cn, pywang@aphy.iphy.ac.cn*)

## Abstract

Using the model for the processive movement of a dimeric kinesin we proposed before, we study the dynamics of a number of mutant homodimeric and heterodimeric kinesins that were constructed by Kaseda *et al.* (Kaseda, K., Higuchi, H. & Hirose, K. (2002) PNAS **99,** 16058–16063). The theoretical results of ATPase rate per head, moving velocity, and stall force of the motors show good agreement with the experimental results by Kaseda *et al.*: The puzzling dynamic behaviors of heterodimeric kinesin that consists of two distinct heads compared with its parent homodimers can be easily explained by using independent ATPase rates of the two heads in our model. We also study the collective kinetic behaviors of kinesins in MT-gliding motility. The results explains well that the average MT-gliding velocity is independent of the number of bound motors and is equal to the moving velocity of a single kinesin relative to MT.

A conventional kinesin is a homodimer with two identical heads linked together by a coiled coil. It converts chemical energy of ATP hydrolysis to mechanical force and can move processively and unidirectionally along a microtubule for hundreds of 8-nm steps without dissociating (1–3). Since its discovery, it has been studied extensively by using various experimental methods and many aspects of its kinetic behavior have been gradually elucidated (4–6).

However, the microscopic mechanism of the processive movement of kinesin is still not well determined. Based on experimental observations, several models have been proposed. One is the thermal ratchet model in which a motor is viewed as a Brownian particle moving in two (or more) periodic but spatially asymmetric stochastically switched potentials (7–9). Another is the "hand-over-hand" model (1, 5, 6, 10–16). In this model, it is supposed that the kinesin motor maintains continuous attachment to the microtubule by alternately repeating single-headed and double-headed binding. This model requires that the two heads move *in a coordinated manner*, which is realized by *a mechanical communication between the two heads during their ATP hydrolysis cycles*. The third model postulates that kinesin head movement is coordinated through an "inchworm" mechanism (6, 17–19), in which also at least one head remains bound to the microtubule during the kinesin movement but the two heads do not swap places.

Recently, we propose another hand-over-hand model which is applicable to both dimeric kinesin (20) and dimeric myosin V (21) that relies on chemical, mechanical, and electrical couplings. In this model, the processive movement *does not require the coordination between the two heads* of the dimeric molecular motors. The dimeric motor steps forward very naturally and, in general, one ATP is hydrolyzed per step (1:1 coupling). This movement of the dimer with 1:1 coupling results solely from that the ATPase rate at the trailing head is much higher than the leading head, caused by the different forces acting on the two heads (the trailing head being pulled forward and the leading head backward). Using the model the dynamics of single kinesin molecules, such as the moving time in one step, the mean moving velocity *versus* [ATP] and load (both positive and negative load), and the stall force *versus* [ATP] are obtained, with results being in good quantitative agreement with previous experimental results (20).

Recently, in order to clarify how each head of kinesin dimer contributes to the stepping movements, Kaseda *et al*. (22) constructed a series of heterodimeric kinesins that consist of two distinct heads (one is WT and the other is mutated). They have observed that ATPase activity of the hybrids was approximately the average of the parent WT and mutant homodimers, but the velocity, stall force, and processivity were not predictable by any simple scheme (23).

Besides the kinetics of single kinesin molecules moving along microtubules, the collective kinetic behaviors of kinesins have also been studied extensively in MT-gliding assays. It is interesting that the MT-gliding velocity is independent of the number of bound kinesin motors (24). Since the motors are not synchronized, this phenomenon is actually not readily understandable.

The purpose of this paper is to use the model we proposed before (20) to study the dynamics of a number of mutant homodimeric and heterodimeric kinesins. Using the model, the quite puzzling experimental results presented by Kaseda *et al.* (22) become easily understandable. The calculated ATPase rate, velocity and stall force for various heterodimers show quite good agreement with the experimental results. The collective behaviors of kinesins in MT-gliding motilities are also studied.

## 1.   Movement Mechanism of Two-Headed Kinesins

The mechanism of processive movement of a two-headed kinesin motor along MT is briefly described as follows (20).

We begin with the dimeric kinesin in rigor state as shown in Fig. 1*a* or Fig. 1*a'*. Kinesin in this state has a larger free energy than the equilibrium (or free) state as shown in Fig. 1*b* or Fig. 1*b'*, which corresponds to the minimum free energy. According to the principle of minimum free energy, the kinesin tends to change from the rigor state to its equilibrium state via an internal elastic force and an internal elastic torque between the two heads.

Let us consider two cases separately:

(i) ATP bound to the trailing head is hydrolyzed earlier. The ATP hydrolysis to ADP and $P_i$ results in the weakening of binding electrostatic force, which is assumed to be resulted from the increase of the relative dielectric constant of the local solvent where ATP hydrolysis has just taken place. Then driven by the internal elastic force and torque, the trailing head detaches from the microtubule and moves to its equilibrium position, as shown in Fig. 1*b*. In this equilibrium state the internal elastic force and torque are zero and there exists only an electrostatic force between the positively-charged detached kinesin head and the neighboring negatively-charged tubulin heterodimer (III). This electrostatic force drives the detached head coming close to heterodimer (III). When the head is close enough to the tubulin heterodimer, the electrostatic force is dependent on the charge distributions on the two surfaces, *i.e.*, the interaction is sterospecific. So that the head will bind the tubulin heterodimer in a fixed orientation, as shown in the figure. From Figs. 1*a* to *c* a mechanochemical cycle is completed, with one ATP being consumed for an effective mechanical step.

(ii) ATP bound to the leading head is hydrolyzed earlier. The leading head then becomes detached. Thus the dimeric kinesin changes from the state in Fig. 1*a'* to the equilibrium state in Fig. 1*b'*. After the original electrical property of the local solvent is recovered the leading head rebinds to the tubulin heterodimer (II). One ATP is hydrolyzed in this futile mechanical cycle.

Now we present the equations used for calculating the ATPase rates of the two kinesin heads. As was pointed out in our previous paper (20) and in Ref. (25), the moving time of a kinesin head in one step (0 ~ 50 μs) is about three orders smaller than the catalytic-turnover time (~ 10 ms) and ATP binding time. The moving velocity of the kinesin motor can thus be considered to be only dependent on the ATPase rates $k$ of the two heads of the motor in rigor state. The ATPase rate of a head depends on ATP concentration [ATP] in a way as described by the Michaelis-Menten kinetics

$$k = \frac{k_c[\text{ATP}]}{[\text{ATP}] + k_c/k_b},\qquad [1]$$

where the catalytic-turnover rate $k_c$ and the ATP binding rate $k_b$ follow the general Boltzman form (26)

$$k_c = \frac{k_c^{(0)}(1+A_c)}{1+A_c\exp(-F\delta/k_BT)},\qquad [2a]$$

$$k_b = \frac{k_b^{(0)}(1+A_b)}{1+A_b\exp(-F\delta/k_BT)}.\qquad [2b]$$

In the above equations, $F$ is the force exerted on the head, $k_c^{(0)}$ and $k_b^{(0)}$ are the corresponding rates at zero force, $\delta$ is a characteristic distance, $A_c$ and $A_b$ are dimensionless constants. When a load $F_{load}$ is acted on the coiled coil that connects the two heads through their neck linkers, the forces on neck linkers of the trailing and leading heads can be written, respectively, as (20)

$$F^T = F_{load} + F_0, \quad [3a]$$

$$F^L = F_{load} - F_0, \quad [3b]$$

where $F_0$ is the internal elastic force. $F_{load}$ is positive when it is plus-end pointed.

## 2. Dynamics of Mutant/Mutant Homodimers and WT/Mutant Heterodimers

In this section we will focus on the study of the dynamics of homodimeric kinesins and of heterodimeric kinesins that consist of two distinct heads, giving explanations to experimental results presented in Ref. (22).

From Eq. **1** it is seen that, at saturating [ATP], the ATPase rate $k$ becomes the same as $k_c$. Therefore, similar to Ref. (22), we will use $k_c$ instead of the ATPase rate $k$ in the following discussions. The catalytic-turnover rates of the two kinesin heads are given by Eqs. **2a**, **3a** and **3b**, with five parameters $\delta$, $A_c$, $k_c^{(0)}$, $F_0$, $F_{load}$. Throughout this paper we will use the fixed values $\delta = 2$ nm and $A_c = 1.1$ for WT and all mutant heads. The value of $A_c$ is the same as that used in Ref. (20) to fit the experimental results of Ref. (27) for WT/WT kinesin. As in experiment (22) we take $F_{load} = 0$. The value of $k_c^{(0)}$ for one head of a homodimer is so taken that the calculated average catalytic-turnover rate per head is in accordance with the experimentally measured value of that homodimer. For the internal elastic force $F_0$, we will make the following assumption: *The internal elastic force $F_0$ for a homodimer is in accordance with the measured stall force of the homodimer*. This assumption is based on the fact that, when $F_{load}$ is equal in magnitude to $F_0$, the net force acted on the neck linker of the trailing head $F^T$ is zero and so the trailing head on average cannot move either toward the forward or toward the backward direction even if it is detached after ATP hydrolysis. The internal elastic force $F_0^{(WT/M)}$ for a WT/mutant heterodimer is produced by the neck linkers of the two distinct heads and can be calculated by using the following equation (see Appendix A)

$$F_0^{(WT/M)} = \frac{2F_0^{WT} F_0^M}{F_0^{WT} + F_0^M}. \quad [4]$$

where $F_0^{WT}$ is the internal elastic force for the WT/WT homodimer and $F_0^M$ is that for the Mutant/Mutant homodimer.

From the experimental values (22) and the above assumption we see that the elastic forces $F_0^{WT}$, $F_0^{L8}$, $F_0^{L11}$, $F_0^{L12}$, $F_0^{L13}$ for WT/WT, L8/L8, L11/L11, L12/L12,

L13/L13, respectively, satisfy the following inequality, $F_0^{WT} > F_0^{L8} > F_0^{L11} > F_0^{L12} > F_0^{L13}$, meaning that the free energy changes $\Delta E$ from free to rigor states of the kinesin homodimers satisfy the following inequality, $\Delta E^{WT} > \Delta E^{L8} > \Delta E^{L11} > \Delta E^{L12} > \Delta E^{L13}$.

In the following, we will calculate the catalytic-turnover rates and stall forces of WT/mutant heterodimers by using the experimental values of mutant homodimers. We also calculate the MT-gliding velocities of all mutant homodimers and heterodimers relative to that of the WT homodimer.

**WT/WT.** $F_0^{WT} = 6$ pN (22). Taking $k_c^{(0)} = 26.9$ s$^{-1}$, the catalytic-turnover rates of the trailing and leading heads calculated by using Eqs. **2a**, **3a** and **3b** with $F_{load} = 0$ are $k_c^T = 53.3$ s$^{-1}$ and $k_c^L = 2.7$ s$^{-1}$, respectively. The total catalytic-turnover rate is $k_c^T + k_c^L = 56$ s$^{-1}$ and the average value per head is $k_c^{(WT/WT)} = 28$ s$^{-1}$, which is thus the same as that determined experimentally (22).

The ratio between the catalytic-turnover rates of the two heads $k_c^T/k_c^L = 19.7$ is much larger than one, and thus the probability that ATP hydrolysis takes place earlier at the leading head than at the trailing head can be neglected. Therefore, the WT/WT generally consumes one ATP to make a forward step (1:1 coupling).

**L8/L8.** $F_0^{L8} = 4$ pN (22). Taking $k_c^{(0)} = 20.2$ s$^{-1}$, the calculated catalytic-turnover rates of the two heads with $F_{load} = 0$ are $k_c^T = 36.6$ s$^{-1}$ and $k_c^L = 4.9$ s$^{-1}$. The total catalytic-turnover rate is $k_c^T + k_c^L = 41.5$ s$^{-1}$ and the average value per head is $k_c^{(L8/L8)} = 20.75$ s$^{-1}$.

The ratio $k_c^T/k_c^L \approx 7.47$ is much larger than one and L8/L8 can be considered to consume one ATP to make a forward step (1:1 coupling). Thus the moving velocity of L8/L8 is about $k_c^{(L8/L8)}/k_c^{(WT/WT)} = 0.74$ times that of WT/WT.

**WT/L8.** Using Eq. **4** we obtain $F_0^{(WT/L8)} = 4.8$ pN. We consider two cases separately. *Case 1*: The WT head is trailing. A force $F^T = 4.8$ pN exerts on the neck linker of the WT head (or it is equivalent to the trailing head of WT/WT with a load $F_{load} = F_0^{(WT/L8)} - F_0^{WT} = -1.2$ pN). Using $k_c^{(0)} = 26.9$ s$^{-1}$ and $F^T = 4.8$ pN for the WT head we get $k_c^T(1) = 51$ s$^{-1}$ from Eq. **2a**. (This value of $k_c^T(1)$ can also be obtained by using parameters for WT/WT homodimer, $F_{load} = -1.2$ pN, and Eqs. **2a** and **3a**.) As the L8 head is leading in this case, a force $F^L = -4.8$ pN exerts on the neck linker of the L8 head (or it is equivalent to the leading head of L8/L8 with a load $F_{load} = F_0^{L8} - F_0^{(WT/L8)} = -0.8$ pN). Using parameters for the L8 head and Eq. **2a** we get $k_c^L(1) = 3.5$ s$^{-1}$. *Case 2*: Similarly, when the L8 head is trailing and the WT head leading, we have $k_c^T(2) = 38.3$ s$^{-1}$ and $k_c^L(2) = 4.6$ s$^{-1}$.

Both ratios $k_c^T(1)/k_c^L(1) \approx 14.6$ and $k_c^T(2)/k_c^L(2) \approx 8.3$ are much larger than one. Thus, as for the case of WT/WT and L8/L8, WT/L8 is also considered to consume one ATP to make a forward step (1:1 coupling). Therefore, the probabilities for occurrences of *Case 1*, *i.e.*, WT as the trailing head, and of *Case 2*, *i.e.*, L8 as the trailing head, are approximately equal and thus the average catalytic-turnover rate per head is $k_c^{(WT/L8)} = \frac{1}{4}\left[k_c^T(1) + k_c^L(1) + k_c^T(2) + k_c^L(2)\right] = 24.35$ s$^{-1}$. The moving velocity of WT/L8 is thus about $k_c^{(WT/L8)}/k_c^{(WT/WT)} = 0.87$ times that of WT/WT.

**L11/L11.** $F_0^{L11} = 1$ pN (22). Taking $k_c^{(0)} = 11.1$ s$^{-1}$ the calculated catalytic-turnover rates of the trailing and leading heads with $F_{load} = 0$ are $k_c^T = 13.9$ s$^{-1}$ and $k_c^L = 8.4$ s$^{-1}$. The total catalytic-turnover rate is $k_c^T + k_c^L = 22.3$ s$^{-1}$ and the average per head is $k_c^{(L11/L11)} = 11.15$ s$^{-1}$.

The ratio $k_c^T/k_c^L = 1.65$ is only slightly larger than one. Thus the probability that ATP hydrolysis takes place earlier at the leading head than at the trailing head cannot be neglected. When ATP hydrolysis takes place earlier at the leading head, the internal elastic force drives it to the equilibrium position, as shown in Fig. 1b'. After the electrical property of the local solution is recovered, the leading head then re-binds to the microtubule. Thus one ATP molecule is consumed in this futile mechanical cycle. When ATP hydrolysis takes place earlier at the trailing head, one ATP molecule is consumed in an effective mechanical cycle. In order to calculate the average number of ATP molecules consumed for an effective mechanical cycle during the kinesin movement, we first determine the probabilities $p_T$ that ATP hydrolysis takes place earlier at the trailing head and $p_L$ that ATP hydrolysis takes place earlier at the leading head. To this end, we take the following approximation: When $k_c^T$ is close to $k_c^L$, $p_T$ is proportional to $k_c^T$ and $p_L$ is proportional to $k_c^L$. Thus $p_T = \frac{k_c^T}{k_c^T + k_c^L}$ and $p_L = \frac{k_c^L}{k_c^T + k_c^L}$. Therefore, on average $\frac{p_T + p_L}{p_T} = \frac{k_c^T/k_c^L + 1}{k_c^T/k_c^L} = 1.6$ molecules of ATP are consumed to make one forward step (1.6:1 coupling). The moving velocity of L11/L11 is $\sim \frac{1}{1.6} \times k_c^{(L11/L11)}/k_c^{(WT/WT)} = 0.25$ times that of WT/WT. It should be pointed that the precise calculations of $p_L$ and $p_L$ need to consider the distributions of the catalytic-turnover time of the two heads, which will be studied in the future.

**WT/L11.** Using Eq. **4** we obtain $F_0^{(WT/L11)} = 1.7$ pN. *Case 1*: The WT head is trailing.

Using $F^T = 1.7$ pN and $k_c^{(0)} = 26.9$ s$^{-1}$ for the WT head we get $k_c^T(1) = 38.1$ s$^{-1}$ from Eq. **2a**. As the L11 head is leading, using $F^L = -1.7$ pN and $k_c^{(0)} = 11.1$ s$^{-1}$ for the L11 head, we get $k_c^L(1) = 6.7$ s$^{-1}$. *Case 2*: Similarly, when the L11 head is trailing and the WT head leading, we have $k_c^T(2) = 15.7$ s$^{-1}$ and $k_c^L(2) = 16.2$ s$^{-1}$.

When the WT (L11) head is trailing (leading), $k_c^T(1)/k_c^L(1) \approx 5.7$ is much larger than one and thus we can consider that one ATP is consumed to make the WT head become the leading head (1:1 coupling). However, when the WT (L11) head becomes leading (trailing), $k_c^T(2)/k_c^L(2) = 0.97$ and thus futile mechanical cycles must be taken into account. For this case we have $p_T(2) \approx 0.5$ and $p_L(2) \approx 0.5$. Thus, on average, $\dfrac{p_T(2) + p_L(2)}{p_T(2)} \approx 2$ molecules of ATP are consumed to make a forward step (2:1 coupling) when L11 (WT) is the trailing (leading) head. In addition, from $p_T(1) \approx 1$, which means that the occurrence of *Case 1* is followed by the occurrence of *Case 2* with a probability of 100%, and from $p_T(2) \approx 0.5$ [$p_L(2) \approx 0.5$], which means that the occurrence of *Case 2* is followed by the occurrence of *Case 1* (*Case 2*) with a probability of 50% (50%), we obtain that the overall probabilities for occurrences of *Case 1* and *Case 2* are 1/3 and 2/3, respectively. The average catalytic-turnover rate per head is thus $k_c^{(WT/L11)} = \dfrac{1}{2} \times \dfrac{1}{3}\left[k_c^T(1) + k_c^L(1)\right] + \dfrac{1}{2} \times \dfrac{2}{3}\left[k_c^T(2) + k_c^L(2)\right] = 18.1$ s$^{-1}$ and the moving velocity of WT/L11 is about $\dfrac{k_c^{(WT/L11)}/k_c^{(WT/WT)}}{(1 \times 1/3 + 2 \times 2/3)} = 0.38$ times that of WT/WT.

**L12/L12.** From the experiment (22) we know that L12 head has a too low affinity for MT so that L12/L12 motor rarely binds to MT. Without the catalysis of MT the ATPase cannot proceed and thus $k_c \approx 0$.

Because the motor rarely binds to MT it is almost always in the free state, as shown in Fig. 2*a*. Because of equal mean electrostatic force in both the forward and the backward directions the motor cannot move unidirectionally.

**WT/L12.** Consider that at one moment the WT head binds strongly to tubulin heterodimer (II) of MT, as shown in Fig. 2*b*. Because of the too low affinity of L12 head for MT, the elastic force, even if it is very small, drives the free L12 head to its equilibrium position, as shown in Fig. 2*b*. At this position, there will be a small net electrostatic force $F_E$ on the L12 head, which has positive charges, pointed to its nearest negatively-charged tubulin heterodimer (III). When ATP hydrolysis takes place at the WT head, the motor then has a larger probability to diffuse forward than backward due to the forward-pointed force $F_E$ and thus the WT head of the WT/L12 motor has a larger probability to bind to the tubulin heterodimer (III). Thus the motor makes a forward step and the motor can move processively to the plus end of MT. From above movement mechanism we see that the stall force, which is the one necessary to stop the forward motion, is equal to $F_E$. From the measured stall force (22) we have $F_E = 0.8$ pN, which is also the force acting on the neck linker of the WT head. Thus using parameter for the WT head we obtain its catalytic-turnover rate as 32.4 s$^{-1}$. As the L12 head has no ATPase activity, the average catalytic-turnover rate per head is $k_c^{(WT/L12)} = 16.2$ s$^{-1}$.

Let us estimate the mechanochemical coupling efficiency. When ATP hydrolysis takes place at the WT head, the probability is $p_+ = J_+/(J_+ + J_-)$ that WT/L12 diffuses forward to a position $x$ at which the net electrostatic force exerted on the WT head by the tubulin heterodimers (I) and (III) becomes strong enough to drive the WT head definitely moving to tubulin heterodimer (III). Here $J_\pm$ denote the drifting rates. The probability is $p_- = J_-/(J_+ + J_-)$ that WT/L12 diffuses backward to the nearly symmetric position $-x$ at which the net electrostatic force becomes strong enough to drive the WT head definitely moving to tubulin heterodimer (I). From the solution of Fokker-Planck equation we have $J_\pm = A\exp(\pm fx/D)$, where $A$ is a constant, $f = F_E/\Gamma$ and $D = k_B T/\Gamma$. From the Stokes formula we obtain $\Gamma = 6\pi\eta r_k = 5.65\times 10^{-11}$ kg s$^{-1}$, where the viscosity $\eta$ of the aqueous medium of a cell around kinesin is approximately $0.01$ g cm$^{-1}$ s$^{-1}$ and the kinesin head is approximated as a sphere with radius $r_k \approx 3$ nm. For calculation we take $x = 1.36$ nm, which corresponds to a net electrostatic force of 4 pN between the WT head and the two heterodimers (I) and (III) along MT, where, to be consistent with that of rat brain kinesin (18), as in Ref. (20) we consider the WT head as a particle with $q_2 = +8$ charges and a tubulin heterodimer as a particle with $q_1 = -27$ charges (28) when the WT head is not very close to a tubulin heterodimer. The vertical distance between the center of the WT head and the negative charge center of tubulin heterodimer is taken as $d_{vertical} = 3.5$ nm (20). Therefore, we obtain that when one ATP is hydrolyzed the WT/L12 has a probability of $p_+ - p_- \approx 0.25$ to move a forward step, *i.e.*, on average, ~ 4 ATP molecules are hydrolyzed to make a forward step (4:1 coupling). The moving velocity of WT/L12 is ~ $\frac{1}{4}\times k_c^{(WT/L12)}/k_c^{(WT/WT)} = 0.145$ of that of WT/WT.

**L13/L13.** Based on experimental results (22, 29), we assume that the internal elastic force for L13/L13 is very weak, *i.e.*, $F_0^{L13} \to 0$, but its affinity for MT is not low. Therefore, even when ATP hydrolysis takes place, the elastic force cannot always overcome the much weakened MT-binding force to drive the motor to its free state. Thus the motor rarely moves. Taking $k_c^{(0)} = 19.8$ s$^{-1}$ and $F_0^{L13} = 0$, from Eq. **2a** we obtain $k_c^{(L13/L13)} = 19.8$ s$^{-1}$.

**WT/L13.** Using Eq. 4 we obtain $F_0^{(WT/L13)} \approx 2F_0^{L13} \to 0$. Thus $F^T \approx 0$, $F^L \approx 0$. Whether trailing or leading, the catalytic-turnover rate of the WT head is 26.9 s$^{-1}$ and that of the L13 head is 19.8 s$^{-1}$. Therefore, the average catalytic-turnover rate per head is $k_c^{(WT/L13)} = 23.35$ s$^{-1}$. Similar to the L13/L13 homodimer, the WT/L13 heterodimer rarely move, too. Since $F_0^{(WT/L13)} \approx 2F_0^{L13} > F_0^{L13}$, WT/L13 may move slightly faster than L13/L13. This is consistent with the experiment (see Table 1 of Ref. (22)).

**WT/E236A.** The E236A head cannot hydrolyze ATP. Using parameters for the WT head, we obtain that, for the elastic force between the two heads in the range of 0 ~ 6 pN, the sum of the catalytic-turnover rates of the WT head when it is trailing and when it is leading is in the range of $53.8 \sim 56$ s$^{-1}$. Therefore, $k_c^{(WT/E236A)} = 13.45 \sim 14$ s$^{-1}$. This is in agreement with the experimental result of $12.9 \pm 1.7$ s$^{-1}$ per head (22).

For clarity, as did in Ref. (22), we summarize the calculated ATPase rates, the MT-gliding velocities, *i.e.*, the moving velocities of single motors (see Sec. 3), and the stall forces of the homodimeric and heterodimeric kinesins in Fig. 3, where all the values are relative to the corresponding values of WT/WT. It is seen that our results show excellent agreement with experimental ones (Fig. 3 of Ref. (22)). This means that our theoretical approach is reliable for studying the dynamics of homodimeric and heterodimeric kinesins.

## 3. Collective Kinetics of Kinesins

In the previous section we focus on the study of dynamics of single heterodimeric kinesins that consist of two distinct heads. In this section we will focus on the study of the cooperative kinetic behaviors of many kinesin motors in MT-gliding motility experiments.

**One Motor.** The statement in Sec. 1 is based on motor movement relative to the MT coordinate, *i.e.*, the MT being fixed. Identically, if the motor is fixed the MT will move toward its minus end with the same velocity. To describe this phenomenon physically, we refer to Fig. 4, where one cycle represents the center-of-mass of the two heads and neck linkers of the motor. The coiled coil which connects the neck linkers to the glass surface is assumed to behave elastically. The state shown by solid lines corresponds to the equilibrium state of the coiled coil (or of the motor-MT system). After ATP hydrolysis and the motor moves a step of *d* to the plus end, *i.e.*, $x = d$, there exists an horizontal elastic force $F$ ($F \equiv -Kx$ with $K$ being the elastic coefficient of the coiled coil) exerting on the motor that binds strongly to the MT via an electrostatic force. Driven by $F$ the motor-MT system moves in over-damped condition toward the MT minus end. When the coiled coil returns to its equilibrium state the MT moves $d$ to the minus end. In another word, when the motor makes, relative to the MT, a step of $d$ toward the plus end the MT will move, relative to the fixed glass coordinate, with the same step length toward the minus end.

**Two Motors.** Now we consider the case of two fixed motors bound to MT. When there is no motor stepping, the system is at equilibrium though the coiled coil of each motor may not be at equilibrium, as shown by solid lines in Fig. 5a, with $x_1 = -a$ and $x_2 = a$. The elastic force exerted by the left coiled coil $F_1 = Ka$ is equal in magnitude to that by the right coiled coil $F_2 = -Ka$, i.e., the net force on the MT in the horizontal direction is $F_{net} = 0$. Then one motor, for example, the left motor makes a step of $d$ on the MT toward the plus end, as shown by dotted lines in Fig. 5a. The two elastic forces, $F_1' = -K(d-a)$ and $F_2 = -Ka$, drive the MT moving $a_{MT}$ toward the minus end and thus the system reaches a new equilibrium state, as shown in Fig. 5b (30). So the position variations of the two motors from the original to present equilibrium states of the system are $\Delta x_1 = d - a_{MT}$ and $\Delta x_2 = -a_{MT}$. As required by equilibrium condition, we have $\Delta F_{net} \equiv -K\Delta x_1 - K\Delta x_2 = 0$, and thus obtain $a_{MT} = d/2$. Now the forces on the two motors are $F_1'' = -K(-a + d - a_{MT}) = -K(d/2 - a)$ and $F_2'' = -K(a - a_{MT}) = -K(a - d/2)$. Similarly, when the right motor makes a step $d$ to the plus end the MT will move $d/2$ to the minus end, too. That means that, for the case of two bound motors, one motor moving a step of $d$ drives the MT moving a step of $d/2$ in the opposite direction.

Using the similar procedure, we deduce that, for the case of three motors bound to the MT, one motor moving a step $d$ drives the MT moving a step $d/3$ in the opposite direction.

**N Motors.** Then we consider the case of $N$ motors bound to the MT in equilibrium state. Assume one motor making a step of $d$. Then the MT has to move in the opposite direction by some distance of $a_{MT}$ to attain the new equilibrium state of the system. This corresponds to that in the fixed glass coordinate the stepping motor actually moves $d - a_{MT}$ toward the plus end of the MT and the other ($N$ – 1) motors move $a_{MT}$ toward the minus end of the MT. As the equilibrium condition requires the net-force change to be zero, i.e., $\Delta F_{net} = -K(d - a_{MT}) + (N-1)Ka_{MT} = 0$, we obtain $a_{MT} = d/N$. In another word, for the case of $N$ bound motors, one motor moving a step of $d$ drives the MT moving $d/N$ in the opposite direction.

Assuming the average ATPase rate of the motors is *k*, we know from the above result that the average MT-gliding velocity with *N* bound motors is $k(d/N) \times N = kd$. This velocity is equal to the average moving velocity of a single motor along a MT, *i.e.*, $kd$. Thus we make a conclusion that, though the motors are unsynchronized mechanochemically, the average MT-gliding velocity is dependent only on the ATPase rate (or moving velocity along MT) of the motors, but not on the number of motors bound to the MT. This is in agreement with the experimental result (Fig. 6 of Ref. (24)).

When there are motors of two different types (thus different ATPase rates), the average MT-gliding velocity should lie between the two velocities as determined by the two different ATPase rates. For simplicity, we discuss with only two motors of different types. Still referring to Fig. 5, assuming the ATPase rate of the left motor is much larger than that of the right motor, then the left motor will tend to make many steps on the MT before the right motor makes a step. Actually, however, the magnitudes of $F_1''$ and $F_2''$ will increase significantly with some continuous stepping of the left motor, and thus the ATPase rates of the left and right motors will be reduced and enhanced, respectively. The final result is that the system arrives at such a state that the two motors have about the same compromised ATPase rate by which the MT-gliding velocity is determined. This should be also true when there are more motors bound to the MT.

It should be noted that the above conclusion is only true when the affinity for MT of the two types of motors are the same. If the motors with a lower ATPase rate have a much lower affinity, they will easily become detached from the MT with the continuous stepping of the higher-ATPase-rate motors. This explains why the MT-gliding velocity with mixed WT/WT and L11/L11 (or L12/L12) homodimers is comparable with the velocity driven by the WT/WT alone (22).

## 4. Conclusion

Using our previous model for processive movement of a dimeric kinesin, 1) we studied the single-molecular dynamic behaviors of a number of mutant homodimeric and heterodimeric kinesins. The puzzling dynamic behaviors of the heterodimeric kinesins that were constructed by Kaseda *et al.* (22) are explained: Our theoretical results for the ATPase rates, moving velocities and stall forces of various homodimeric and heterodimeric kinesins show quite good agreement with the experimental ones. 2) We studied the collective kinetic behaviors of kinesins. Our theoretical results explain well the experimental results that the average MT-gliding velocity is independent of the number of bound kinesin motors although they are not synchronized and is equal to the moving velocity of a single kinesin relative to the MT.

The project is supported by the National Natural Science Foundation of China (Grant Numbers 60025516, 10334100).

## Appendix A

Assume that the elastic coefficients for the neck linkers of a WT head and a mutant head are $K^{WT}$ and $K^M$, respectively. The effective elastic coefficient $K^{(WT/M)}$ of the two neck linkers which connect the WT head and the mutant head in the WT/Mutant heterodimer can be obtained by considering it as a series connection of the neck linker of the WT head and that of the mutant head. If the total length change of the neck linkers of the WT/Mutant heterodimer is $\Delta l$, the length change of the WT neck linker is $x^{WT}$, and that of the mutant neck linker is $x^M$, then we should have

$$x^{WT} + x^M = \Delta l, \qquad [A1]$$

$$K^{WT} x^{WT} + K^M x^M = K^{(WT/M)} \Delta l, \qquad [A2]$$

$$K^{WT} x^{WT} = K^M x^M. \qquad [A3]$$

Solving Eqs. **A1-A3**, we can easily obtain

$$K^{(WT/M)} = \frac{2 K^{WT} K^M}{K^{WT} + K^M}. \qquad [A4]$$

Writing Eq. **A4** in the form of elastic force, we get Eq. **4**.

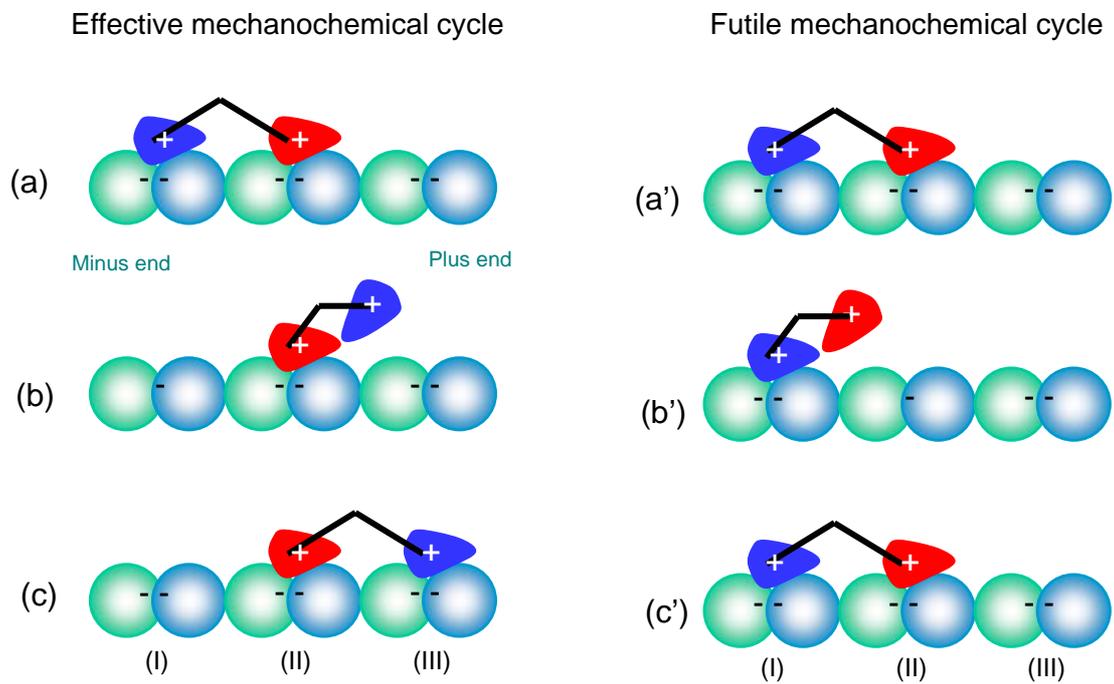

**Fig. 1.** Schematic illustrations of kinesin movement mechanism. The two kinesin heads are in blue and red. The α- and β-tubulin subunits from a single microtubule protofilament are indicated in light green and light blue. Each pair of α- and β-tubulin subunits forms a tubulin heterodimer. The polarity of the microtubule is indicated. **Effective mechanochemical cycle:** (a) The cycle begins with both heads binding to the microtubule. Note that the elastic force exerts on the two heads in opposite directions. (b) ATP hydrolysis at the blue head (trailing head) changes the electrical property of the local solution, causing an increase of the dielectric constant of the local solution. This in turn weakens the electrostatic force between this head and the tubulin heterodimer (I), resulting in the detachment and subsequent movement of the blue head towards a position as determined by the equilibrium structure of the kinesin dimer. (c) The blue head binds to the new tubulin heterodimer (III) via the electrostatic interactions and becomes the leading head for the next cycle. One ATP is hydrolyzed for this 8-nm forward step. **Futile mechanochemical cycle:** (a') The cycle also begins with both heads binding to the microtubule. (b') ATP hydrolysis at the red head (leading head) weakens the electrostatic force between this head and the tubulin heterodimer (II), resulting in the detachment and subsequent movement of the red head towards the equilibrium position. (c') The red head rebinds to the tubulin heterodimer (II) after the original electrical property of local environment of the solution is recovered. One ATP is hydrolyzed in this futile mechanical cycle.

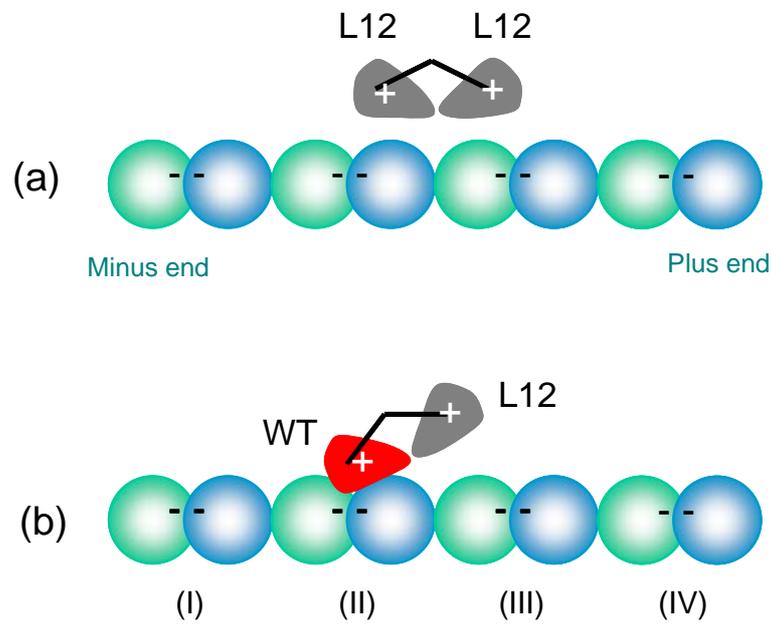

**Fig. 2.** (a) The L12/L12 does not bind to MT and is in the free state. (b) In the rigor state, the WT head of motor WT/L12 binds strongly to a tubulin heterodimer of MT, and because of the too low affinity of L12 head for MT, the elastic force drives the free L12 head to its equilibrium position.

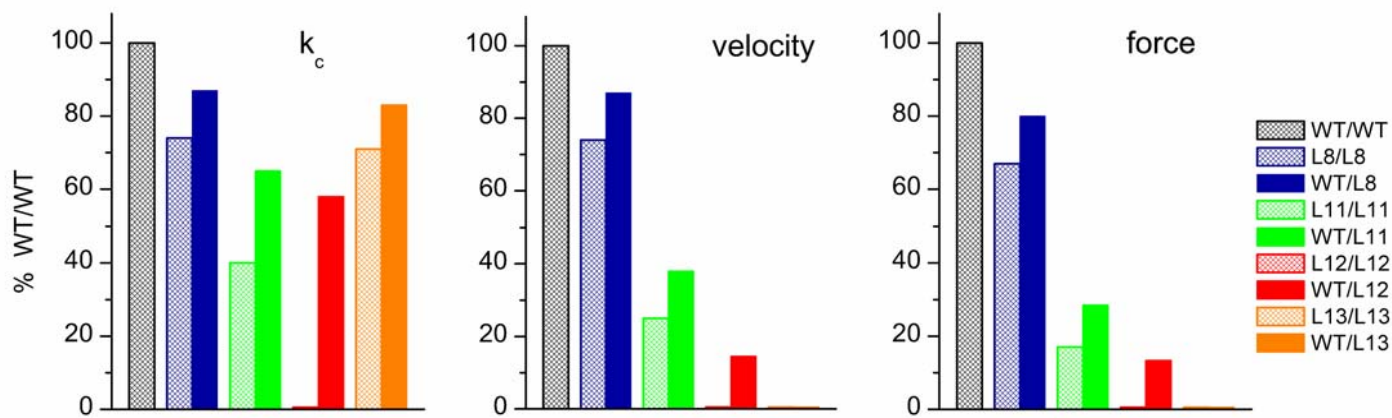

**Fig. 3.** The calculated ATPase rates, MT-gliding velocities, and stall forces of homodimeric and heterodimeric kinesins relative to that of WT/WT homodimer.

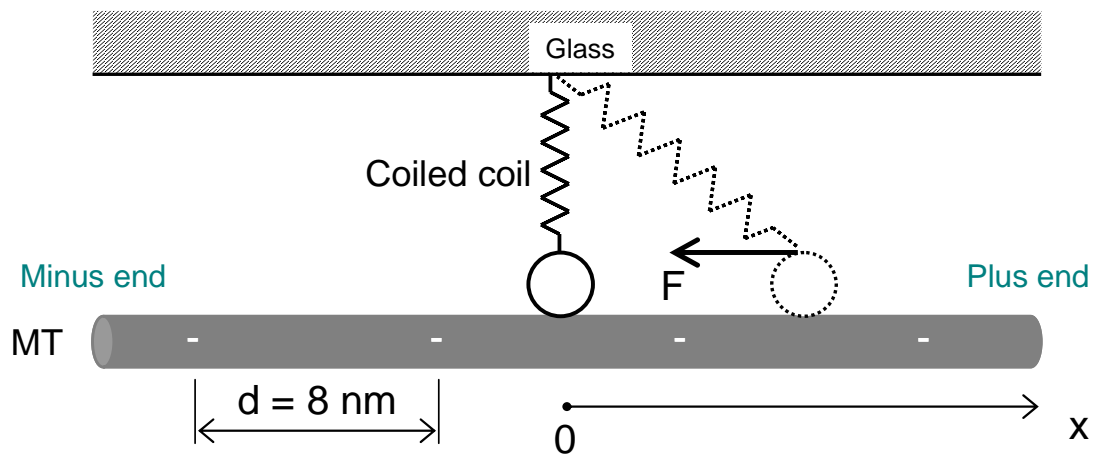

**Fig. 4.** Geometry of the MT gliding with a single kinesin homodimer. The center-of-mass of the two heads and two neck linkers of the motor are represented by a circle, and the coiled coil which connects the neck linkers to the glass surface is represented by a zigzag line. The symbols "-" represent the binding sites on MT for kinesin heads. The state shown by solid lines corresponds to the equilibrium state of the coiled coil (or of the motor-MT system), and the state by dotted lines to that the motor moves a step of *d* to the plus end of MT after ATP hydrolysis. Note that the center-of-mass of the motor is at the middle position between two neighboring binding sites.

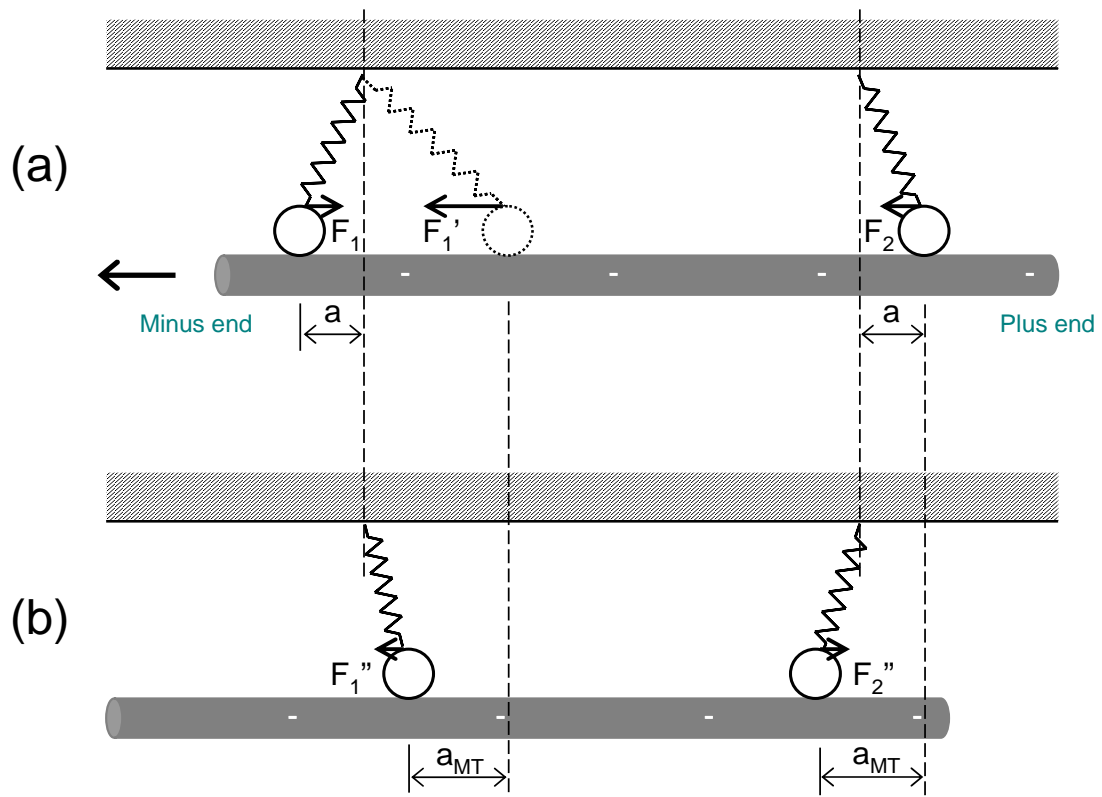

**Fig. 5.** Geometry of the MT gliding with two kinesin homodimers.